\documentclass{cimento}

\usepackage{amssymb}
\usepackage{lineno}
\usepackage{graphicx}

\def\NIMA{Nucl. Instrum. Methods A}

\def\PLB{Phys. Lett.  B}

\def\PRD{Phys. Rev. D}

\def\EPJC{Eur. Phys. J. C}
\def\CHPC{Chin. Phys. C}

\def\be{\begin{equation}}
\def\ee{\end{equation}}
\def\bea{\begin{eqnarray}}
\def\eea{\end{eqnarray}}

\newcommand{\kpimmws}{K^{\pm} \to \pi^{\mp} \mu^{\pm} \mu^{\pm}}
\newcommand{\kpimmns}[1]{K^{#1} \to \pi \mu \mu}
\newcommand{\kpimm}[1]{K^{#1} \to \pi^{#1} \mu^{+} \mu^{-}}

\newcommand{\kmutwo}[1]{K^{#1} \to \mu^{#1} \nu}
\newcommand{\kltwo}[1]{K^{#1} \to \ell^{#1} \nu}
\newcommand{\kmutwoN}[1]{K^{#1} \to \mu^{#1} N_{4}}

\newcommand{\Npimuws}{N_{4} \to \pi^{\mp}\mu^{\pm}}
\newcommand{\Npimurs}{N_{4} \to \pi^{\pm}\mu^{\mp}}
\newcommand{\Npimuns}{N_{4} \to \pi\mu}

\newcommand{\kthreepic}[1]{K^{#1} \to \pi^{#1} \pi^+ \pi^-}
\newcommand{\kthreepin}[1]{K^{#1} \to \pi^{#1} \pi^0 \pi^0}
\newcommand{\kpmm}{K_{\pi\mu\mu}}
\newcommand{\kpmmlnv}{K_{\pi\mu\mu}^{\rm LNV}}
\newcommand{\kpmmlnc}{K_{\pi\mu\mu}^{\rm LNC}}
\newcommand{\npmmlnv}{N_{\pi\mu\mu}^{\rm LNV}}

\newcommand{\eqdef}{=}
\newcommand{\ttiny}[1]{\texttt{\tiny #1}}

\title{New limits on heavy neutrinos from Kaon experiments at CERN}

\author{K.~Massri\thanks{for the NA48/2 Collaboration: Cambridge, CERN, Dubna, Chicago, Edinburgh, Ferrara, Firenze, Mainz, Northwestern, Perugia, Pisa, Saclay, Siegen, Torino, Wien,\\
and the NA62-2007 collaboration: Birmingham, CERN, Dubna, Fairfax, Ferrara, Firenze, Frascati, Mainz, Merced, Moscow, Napoli, Perugia, Pisa, Protvino, Roma I, Roma II, Saclay, San Luis Potosi, Sofia, Stanford, Torino, TRIUMF.}}
\instlist{\inst{CERN - Geneva, Switzerland}}

\begin{document}
\maketitle

\begin{abstract}
The NA48/2 and NA62-$R_K$ experiments at CERN collected large samples of charged kaon decays in 2003--2004 and 2007, respectively.
These samples, collected with different trigger conditions, allow to search for both short and long\mbox{-}living heavy neutrinos produced in $\kmutwoN{\pm}$ decays. The results of these complementary searches are presented in this letter.
In the absence of observed signal, the limits obtained on $\mathcal{B}(\kpimmws)$, $\mathcal{B}(K^{\pm}\to \mu^{\pm} N_4)\mathcal{B}(N_4\to \pi\mu)$, $\mathcal{B}(\kmutwoN{+})$ and on the mixing matrix element $|U_{\mu4}|^2$ are reported.
\end{abstract}


\section{Introduction}
Neutrinos are strictly massless within the Standard Model~(SM), due to the absence of right-handed neutrino states. However, since the observation of neutrino oscillations~\cite{pdg} has unambiguously demonstrated the massive nature of neutrinos, right-handed neutrino states must be included.
A natural extension of the SM involves the inclusion of $n_s$ sterile neutrinos which mix with ordinary neutrinos to explain several open questions. An example of such a theory is the Neutrino Minimal Standard Model~($\nu$MSM)~\cite{as05_02}.
In this model, three massive right-handed neutrinos are introduced to explain simultaneously neutrino oscillations, dark matter and baryon asymmetry of the Universe: the lightest has mass~$\mathcal{O}(1\mbox{ keV})$ and is a dark matter candidate; the other two, with masses ranging from 100~MeV/$c^2$ to few GeV/$c^2$, are responsible for the masses of the SM neutrinos (via see-saw mechanism) and introduce extra CP violating phases to account for baryon asymmetry.
These SM extensions predict new particles, such as heavy neutrinos~$N_i$~($i = 4,\dots,n_s+3$), which could be produced in kaon decays\footnote{For simplicity, only the case $n_s=1$ will be considered onwards.} and, depending on their lifetime, possibly decay into visible final states.

The NA48/2 and NA62-$R_K$ experiments at CERN collected large samples of charged kaon decays in 2003--2004 and 2007, respectively.
These samples, collected with different trigger conditions, allow to search for both short and long-living
heavy neutrinos produced in $\kmutwoN{\pm}$ decays.
In particular, short-living heavy neutrinos decaying promptly to $\Npimuns$, as well as off-shell Majorana neutrinos mediating Lepton Number Violating (LNV) processes, can be searched in the $\kpimmns{\pm}$ samples collected by the NA48/2 experiment~\cite{ba17}, while long-living heavy neutrinos escaping detection can be searched by looking for peaks in the missing mass spectrum of the $\kmutwo{\pm}$ candidates collected by the NA62-$R_K$ experiment~\cite{la17}.
The results of these complementary searches are presented in this letter.

\section{Experimental Apparatus and Data Taking conditions}
\label{sec:detector}
The NA48/2 experiment at CERN SPS was a multi-purpose $K^{\pm}$ experiment which collected data in 2003--2004, whose main goal was to search for direct CP violation in the $\kthreepic{\pm}$ and $\kthreepin{\pm}$ decays~\cite{ba07}.
Simultaneous and collinear $K^+$ and $K^-$ beams of the same momentum ($60\pm3.7$)~GeV/$c$ were produced by the 400~GeV/$c$ SPS primary proton beam, which impinged on a Beryllium target, and were steered into a 114~m long decay region, contained in a vacuum (at pressure $< 10^{-4}$~mbar) cylindrical tank.
The downstream part of the vacuum tank was sealed by a convex Kevlar window, that separated the vacuum from the helium at atmospheric pressure in which a magnetic spectrometer, formed of 4 drift chambers (DCHs) and a dipole magnet providing a horizontal momentum kick~$p_t = 120$~MeV$/c$, was installed.
The spatial resolution of each DCH was $\sigma_x = \sigma_y = 90$~$\mu$m, while the momentum resolution of the spectrometer was~$\sigma(p)/p = (1.02 \oplus 0.044 \cdot p)\%$, where the momentum~$p$ is measured in GeV/$c$.
A hodoscope~(HOD)
was placed downstream of the spectrometer and provided fast signals for trigger purposes, as well as time measurements for charged particles with a resolution of $\sim 300$~ps.
The HOD was followed by a LKr electromagnetic calorimeter
with a depth of 127~cm, corresponding to 27 radiation lengths.
The front plane had an octagonal shape and was segmented in 13248 cells with size $2 \times 2$~cm$^2$.
The LKr calorimeter energy resolution was measured to be~$\sigma_E/E = (3.2/\sqrt{E} \oplus 9.0/E \oplus 0.42)\%$,
where $E$ is the energy expressed in GeV. The space resolution~$\sigma_{x,y}$ of the LKr was~$\sigma_{x,y} = (4.2/\sqrt{E} \oplus 0.6)~\mbox{mm}$, and the time resolution on the single shower was $\sigma_t = 2.5\mbox{ ns}/\sqrt{E}$.
The LKr was followed by a hadronic calorimeter (not used for the present measurement) and a muon detector~(MUV). The MUV consisted of three $2.7\times2.7$~m$^2$ planes of plastic scintillator strips, each preceded by a 80~cm thick iron wall and alternately aligned horizontally and vertically.
The strips were 2.7~m long and 2~cm thick, and they were read out by photomultipliers at both ends. The widths of the strips were 25~cm in the first two planes, and 45~cm in the third plane. A detailed description of the NA48/2 beam line and the detector layout can be found in Refs.~\cite{ba07,fa07}.

The NA62 experiment ($R_K$ phase, denoted as NA62-$R_K$), whose main goal was to measure the ratio $R_K$ of the rates of the $\kltwo{\pm}$ decays ($\ell = e,\mu$), collected a large minimum bias data sample in 2007-2008~\cite{la13}. It was essentially based on the NA48/2 detector with minor changes. The beam momentum was changed to ($74\pm1.4$)~GeV/$c$ and the spectrometer momentum kick was increased to~$p_t = 270 $~MeV/$c$, which led to an improved momentum resolution~$\sigma(p)/p = (0.48 \oplus 0.009 \cdot p)\%$. In addition, the trigger system was modified to collect single track leptonic modes.

\section{Search for off-shell and short-living heavy neutrinos at NA48/2}
Searches for the LNV $\kpimmws$ decay, which is forbidden in the SM and could proceed via off-shell Majorana neutrinos, and for short-living heavy neutrinos decaying promptly to $\Npimuns$ are performed in the $\kpimmns{\pm}$ samples collected by the NA48/2 experiment~\cite{ba17}. 
Since a heavy neutrino~$N_4$ produced in a $K^{\pm}\to\mu^{\pm}N_4$ decay and decaying promptly to $\pi^{\pm}\mu^{\mp}$ would produce a narrow spike in the invariant mass $M_{\pi\mu}$ spectrum, the invariant mass distributions of the collected $\kpimmns{\pm}$ samples have been scanned looking for such a signature. 

\boldmath
\subsection{Selected data samples}
\unboldmath
\label{sec:kpimmws_datasamples}
The event selection is based on the reconstruction of a three-track vertex: given the resolution of the vertex longitudinal position ($\sigma_{vtx} = 50$~cm), $\kpimmws$ and $\kpimm{\pm}$ decays (denoted $\kpmmlnv$ and $\kpmmlnc$ below) mediated by a short-lived ($\tau\lesssim 10$~ps) resonant particle are indistinguishable from a genuine three-track decay.
The size of the selected $\kpmm$ samples is normalised relative to the abundant $K^\pm\to\pi^\pm\pi^+\pi^-$ channel (denoted $K_{3\pi}$ below),
from which the number of $K^\pm$ decays in the 98~m long fiducial decay region is obtained: $N_K = (1.637 \pm 0.007)\times 10^{11}$.
The $\kpmm$ and $K_{3\pi}$ samples are collected concurrently using the same trigger logic.

\begin{figure}[h]
\begin{minipage}{0.5\textwidth}
\includegraphics[width=\textwidth]{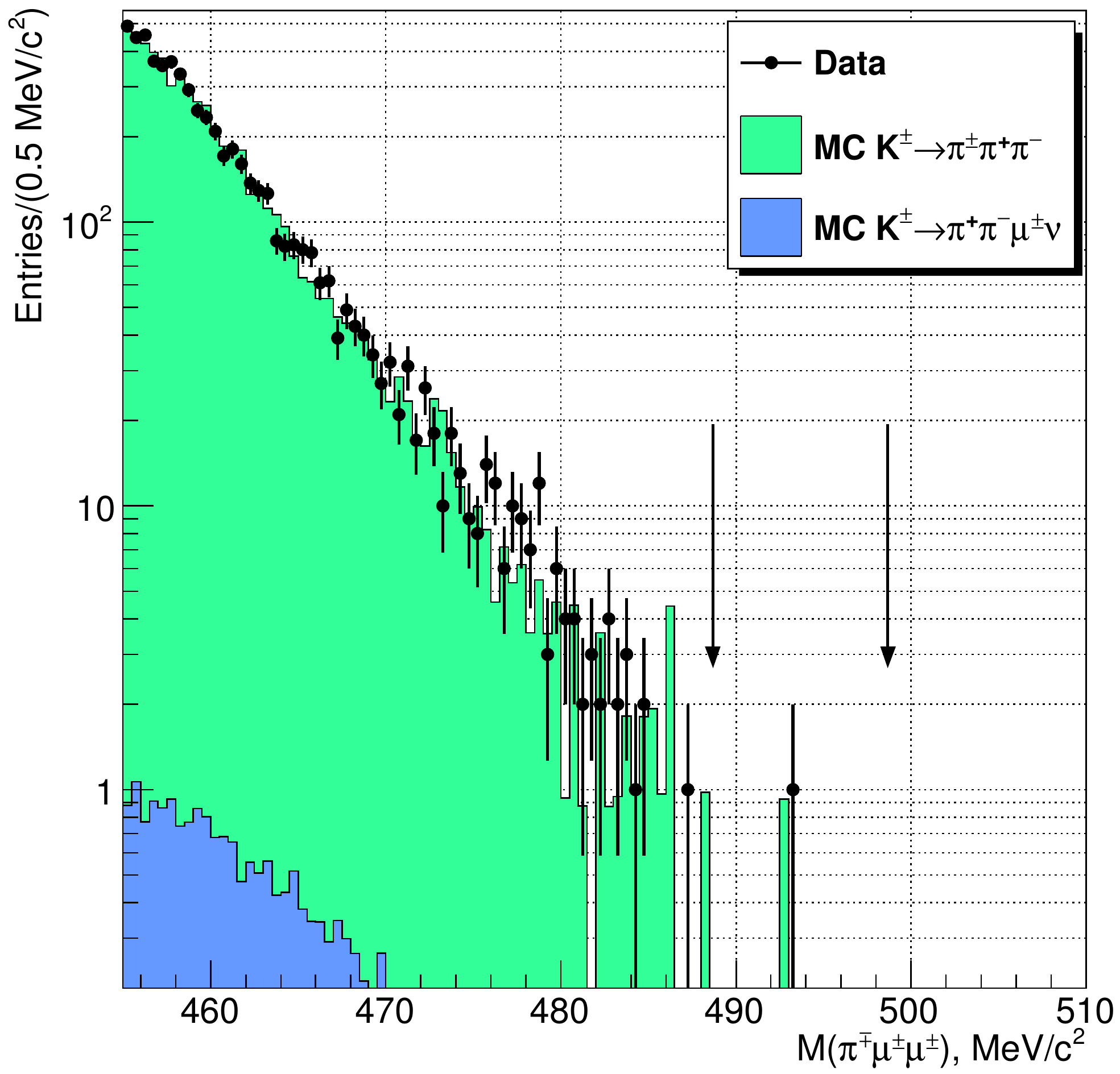}%
\end{minipage}
\hfill
\begin{minipage}{0.5\textwidth}
\includegraphics[width=\textwidth]{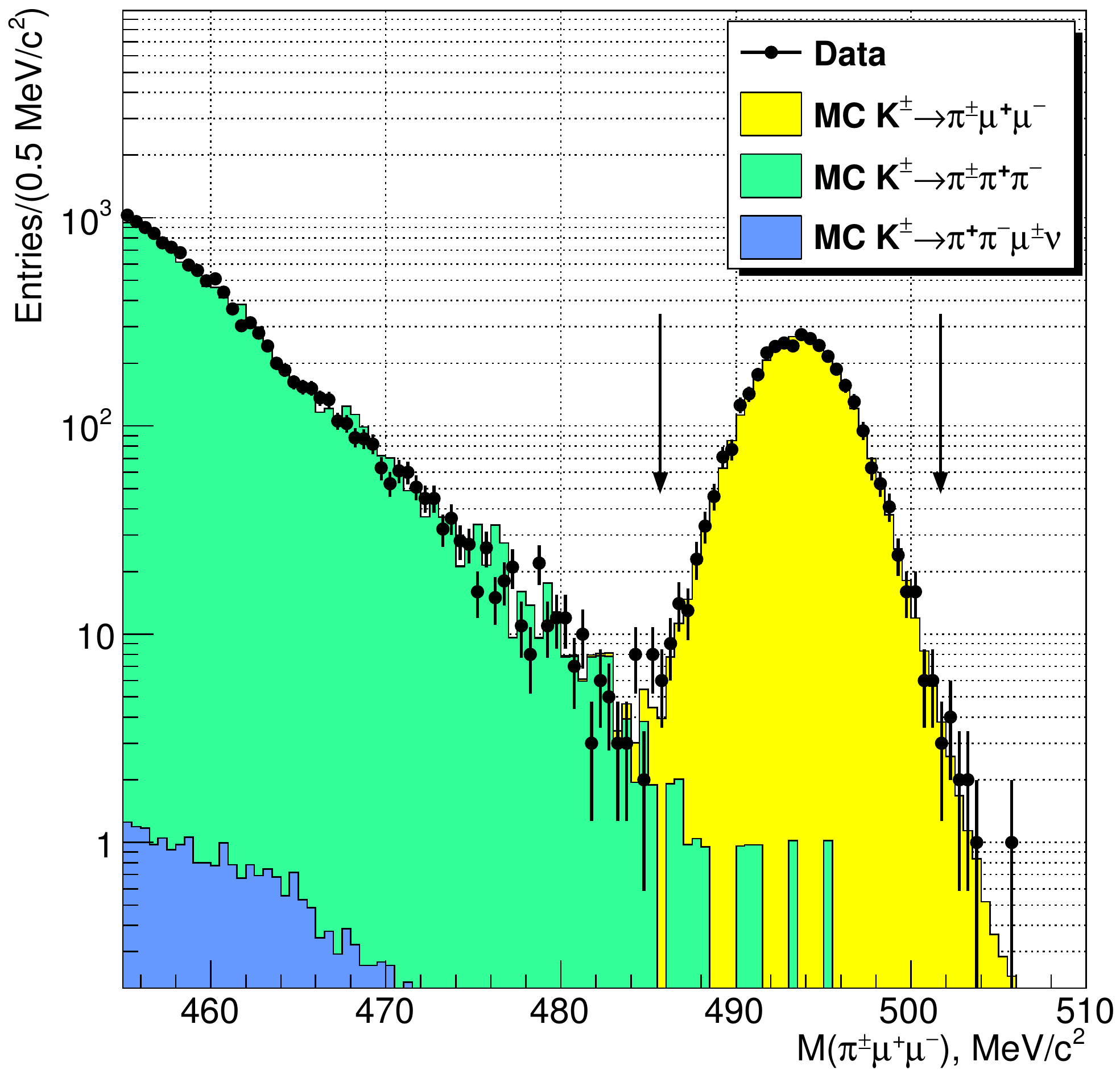}%
\end{minipage}
\caption{Invariant mass distributions of data and MC events passing the $\kpmmlnv$ (left) and
$\kpmmlnc$~(right) selections. The signal mass regions are indicated with vertical arrows.}\label{fig:mpimm}  
\end{figure}

The invariant mass distributions of data and MC events passing the $\kpmmlnv$ and $\kpmmlnc$ selections are shown in Fig.~\ref{fig:mpimm}.
One event is observed in the signal region after applying the $\kpmmlnv$ selection, while 3489 $\kpmmlnc$ candidates are selected with the $\kpmmlnc$ selection. A peak search assuming different mass hypotheses is performed over the distributions of the invariant masses $M_{\pi\mu}$ of the selected $\kpmm$ samples. In total, 284 (267) mass hypotheses are tested respectively for the search of resonances in the $M_{\pi\mu}$ distribution of the $\kpmmlnv$ ($\kpmmlnc$) candidates, covering the full kinematic ranges.

\boldmath
\subsection{Upper Limit on $\mathcal{B}(\kpimmws)$}
\unboldmath
The upper limit~(UL) at 90\% confidence level~(CL) on the number of $\kpimmws$ signal events in the $\kpmmlnv$ sample corresponding
to the observation of one data event and a total number of expected background events $N_{bkg} = 1.160 \!\pm\! 0.865$ is obtained applying an extension of the Rolke-Lopez method~\cite{ro01}: $\npmmlnv < 2.92$ at 90\%~CL. 
Using the values of the signal acceptance~$A(\kpmmlnv)=20.62\%$ estimated with MC simulations and the number~$N_K$ of kaon decays in the fiducial volume~(Sec.~\ref{sec:kpimmws_datasamples}), the UL on the number of $\kpimmws$ signal events in the $\kpmmlnv$ sample leads to a constraint on the signal branching ratio~$\mathcal{B}(\kpimmws)$:
\begin{equation}
\label{eq:BR_kpimmws_experimental}
\mathcal{B}(\kpimmws) = \frac{\npmmlnv}{N_K\cdot A(\kpmmlnv)}< 8.6 \times 10^{-11} \quad \mbox{@ 90\% CL}.
\end{equation}

\subsection{Limits on short-living heavy neutrino decay}
\begin{figure}[h]
\begin{center}
\begin{minipage}{0.5\textwidth}
\includegraphics[width=\textwidth]{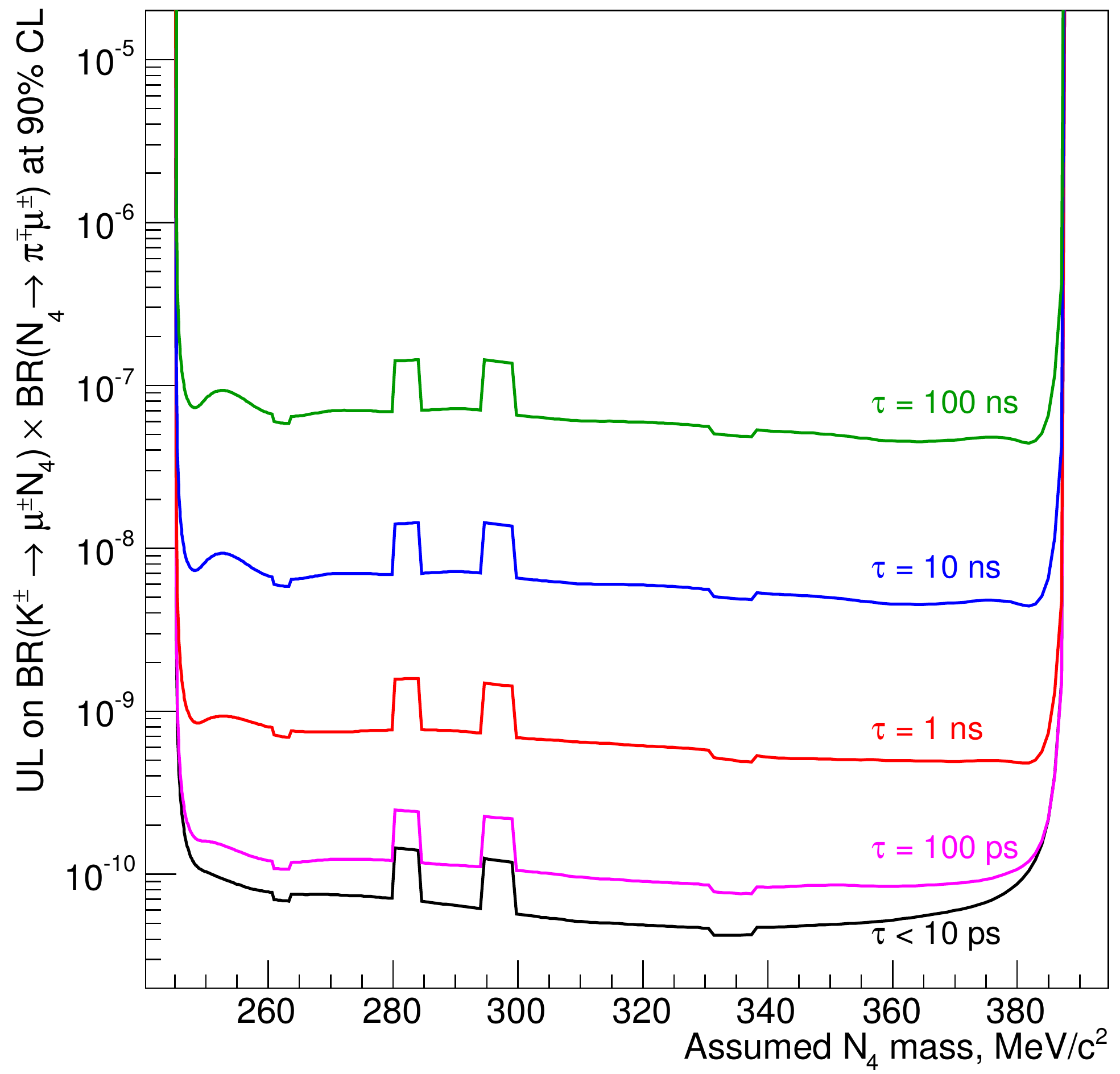}%
\end{minipage}
\put(-23,80){\Large\bf a}
\hfill
\begin{minipage}{0.5\textwidth}
\includegraphics[width=\textwidth]{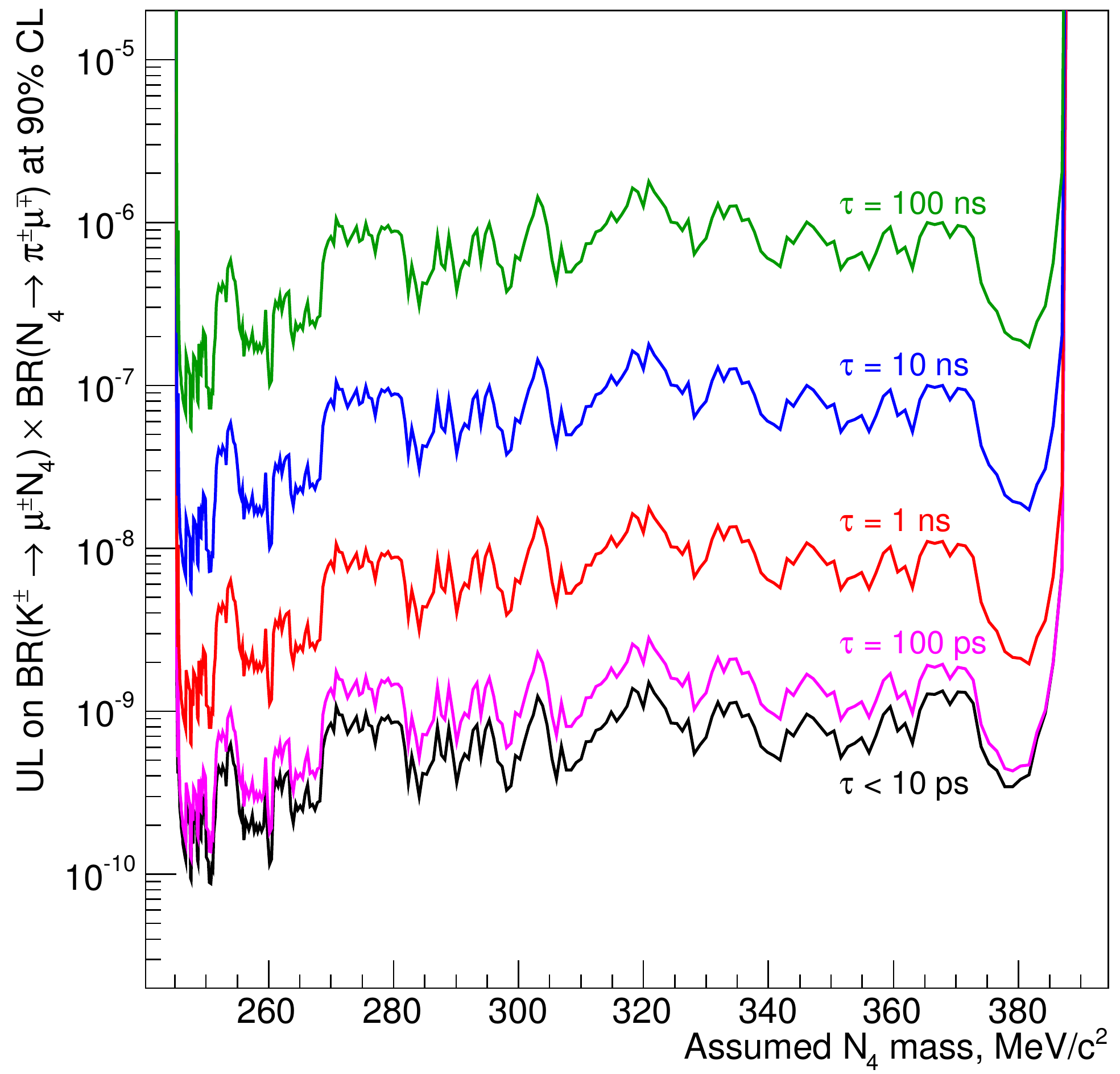}%
\end{minipage}
\put(-23,80){\Large\bf b}
\caption{Obtained ULs at 90\% CL on the products of branching ratios as functions of the resonance mass and lifetime: a)~$\mathcal{B}(\kmutwoN{\pm})\mathcal{B}(\Npimuws)$; b)~$\mathcal{B}(\kmutwoN{\pm})\mathcal{B}(\Npimurs)$.
All presented quantities are strongly correlated for neighbouring resonance masses as the mass step of the scan is about 8 times smaller than the signal window width.}\label{fig:kpimmws_results_BR}  
\end{center}
\end{figure}

\begin{figure}[h]
\begin{center}
\begin{minipage}{0.5\textwidth}
\includegraphics[width=\textwidth]{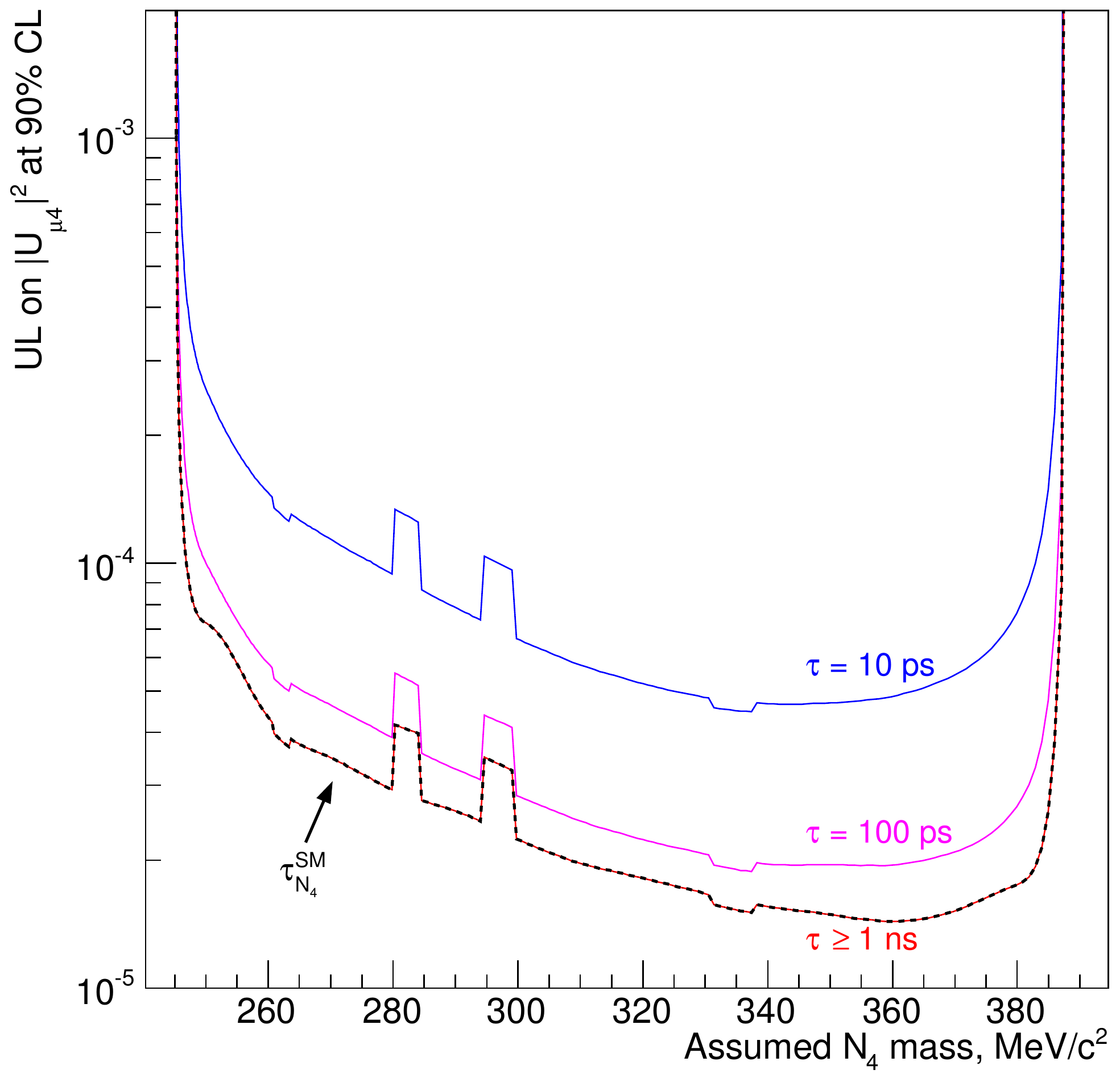}%
\end{minipage}
\put(-23,80){\Large\bf a}
\hfill
\begin{minipage}{0.5\textwidth}
\includegraphics[width=\textwidth]{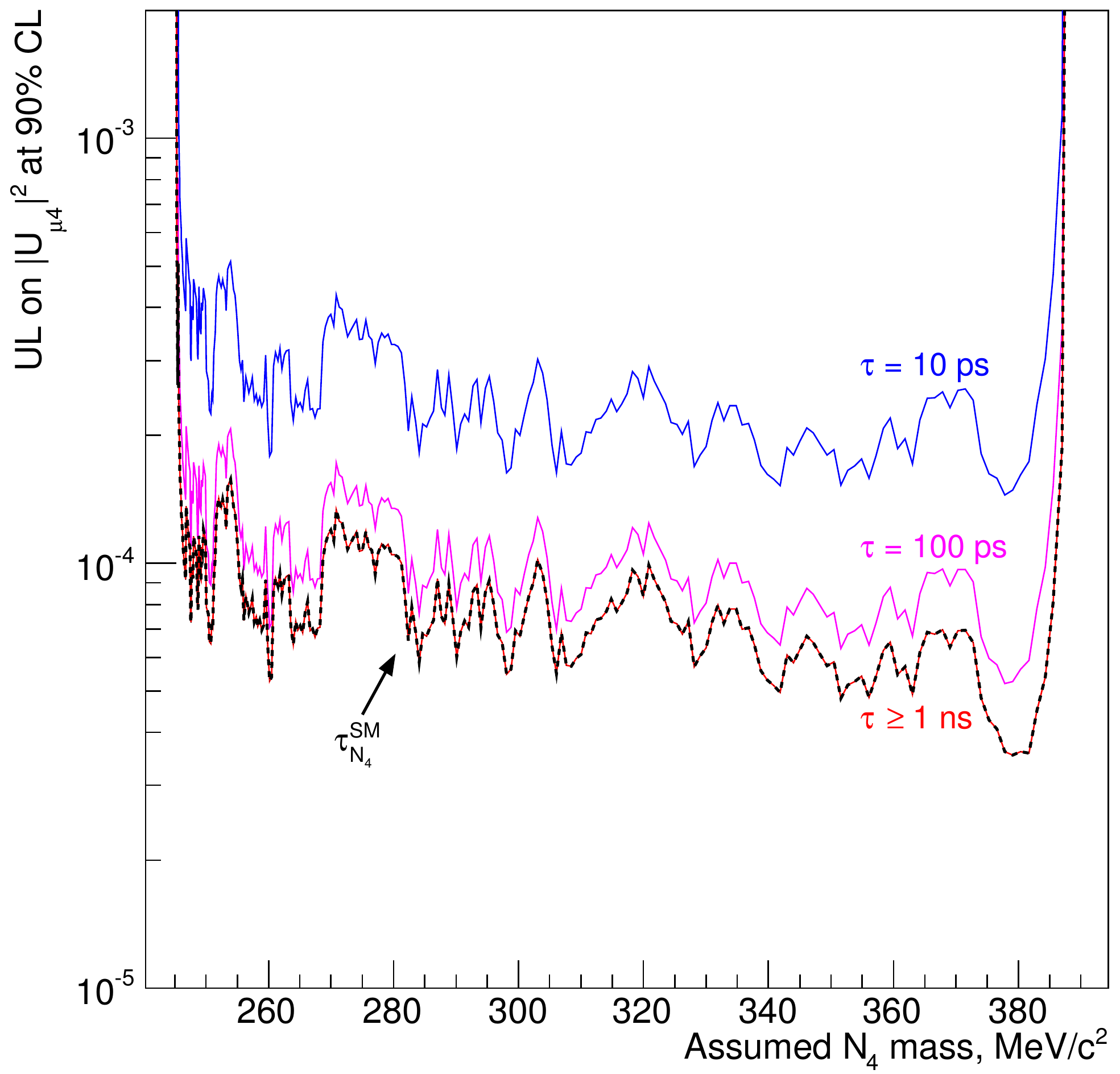}%
\end{minipage}
\put(-23,80){\Large\bf b}
\caption{Upper limits at 90\% CL on $|U_{\mu 4}|^2$ as functions of the assumed resonance mass and lifetime obtained from the limits on: a)~$\mathcal{B}(\kmutwoN{\pm})\mathcal{B}(\Npimuws)$; b)~${\mathcal{B}(\kmutwoN{\pm})\mathcal{B}(\Npimurs)}$. The boundaries for $\tau \geq 1$~ns are valid up to a maximum lifetime of $\sim 100$~$\mu$s.}\label{fig:kpimmws_results_coupling}
\end{center}
\end{figure}

No signal is observed, as the local significances of the signals in each mass hypothesis never exceed 3 standard deviations.
In absence of a signal, ULs on the product~$\mathcal{B}(K^{\pm}\to \mu^{\pm} N_4)\mathcal{B}(N_4\to \pi\mu)$ as a function of the resonance lifetime~$\tau$ are obtained for each mass hypothesis~$M_i$, by using the values of the acceptances~$A_{\pi\mu\mu}(M_i,\tau)$ and the ULs on the number~$N^i_{sig}$ of signal events for such a mass hypothesis:
\begin{equation}
\label{eq:BR_res_massbin}
\left.\mathcal{B}(K^{\pm}\to \mu^{\pm}N_4)\mathcal{B}(N_4\to \pi\mu)\right|_{M_i,\tau} = \frac{N_{sig}^i}{N_K \cdot A_{\pi\mu\mu}(M_i,\tau)}.
\end{equation}
The obtained ULs on~$\mathcal{B}(K^{\pm}\to \mu^{\pm}N_4)\mathcal{B}(N_4\to \pi\mu)$ as a function of the resonance mass, for several values of the resonance lifetime, are shown in Fig.~\ref{fig:kpimmws_results_BR}.
Limits on the products~$\mathcal{B}(\kmutwoN{\pm})\mathcal{B}(\Npimuns)$ obtained from $\kpmmlnv$ and $\kpmmlnc$ samples can be
used to constrain the squared magnitude $|U_{\mu4}|^2$ using the relation~\cite{cv10}
$$
|U_{\mu4}|^2 = \frac{8\sqrt{2}\pi\hbar}{G_F^2 \sqrt{M_K \tau_K} f_Kf_{\pi}|V_{us}V_{ud}|} \sqrt{\frac{\mathcal{B}(\kmutwoN{\pm})\mathcal{B}(\Npimuns)}{\tau_{N_4} M_{N_4}^5\lambda^{\frac{1}{2}}(1,r_{\mu}^2,r_{N_4}^2)\lambda^{\frac{1}{2}}\left(1,\rho_{\pi}^2,\rho_{\mu}^2\right)\chi_{\mu\mu}}},
$$
where $r_{i} \eqdef M_{i}/M_K$, $\rho_i \eqdef M_{i}/M_{N_4}$ ($i=\mu,\pi, N_4$), $\lambda(a,b,c) \eqdef a^2 + b^2 + c^2 -2ab -2ac -2bc$ and $\chi_{\mu\mu} = [(1+\rho_{\mu}^2)-(r_{N_4}^2-r_{\mu}^2)(1-\rho_{\mu}^2)][(1-\rho_{\mu}^2)^2-(1+\rho_{\mu}^2)\rho_{\pi}^2]$.
The value of the lifetime~$\tau_{N_4}^{\ttiny{SM}}$, obtained assuming that the heavy neutrino decays into SM particles only and that $|U_{e4}|^2 = |U_{\mu4}|^2 = |U_{\tau4}|^2$, is evaluated for each mass hypothesis, using the decay widths provided in Ref.~\cite{at09}.
The ULs on $|U_{\mu4}|^2$ as functions of the resonance mass obtained for several values of the assumed resonance lifetime, including~$\tau_{N_4}^{\ttiny{SM}}$, are shown in Fig.~\ref{fig:kpimmws_results_coupling}.

\boldmath
\section{Search for long-living heavy neutrinos at NA62-$R_K$}
\unboldmath
A search for long-living heavy neutrinos with masses in the range 300--375 MeV/$c^2$ is performed in the $\kmutwo{+}$ sample collected by the NA62-$R_K$ experiment~\cite{la17}.
Since a heavy neutrino~$N_4$ produced in a $\kmutwoN{+}$ decay would produce a narrow spike in the missing mass $M_{miss} = \sqrt{(p_K - p_{\mu})^2}$ spectrum, where $p_K$ and $p_{\mu}$ are the kaon and muon four-momenta respectively, the $M_{miss}$ distribution of the $\kmutwo{+}$ candidates has been scanned looking for such a signature.
The considered mass range is constrained by the existing strong limits on $|U_{\mu4}|^2$ set up to 300 MeV/$c^2$ by the BNL-E949 experiment~\cite{ar15} and by the drop on heavy neutrino acceptance above 375 MeV/$c^2$, while the choice of considering data with the $K^+$ beam only is dictated by the muon halo background being smaller in the $K^+$ sample.
The heavy neutrinos are assumed to decay only to SM particles and have a lifetime $\tau \gtrsim 1$~$\mu$s (which corresponds to $|U_{\mu4}|^2 < 10^{-4}$), such that their mean free path in the mass range considered is longer than 10 km. In this case, their decay can then be neglected as the probability of decaying within the detector is below 1\%.
Since only the production process is studied, the limits scale linearly with the kaon flux.

\boldmath
\subsection{Selected data samples}
\begin{figure}[h]
\begin{minipage}{0.517\textwidth}
\includegraphics[width=\textwidth]{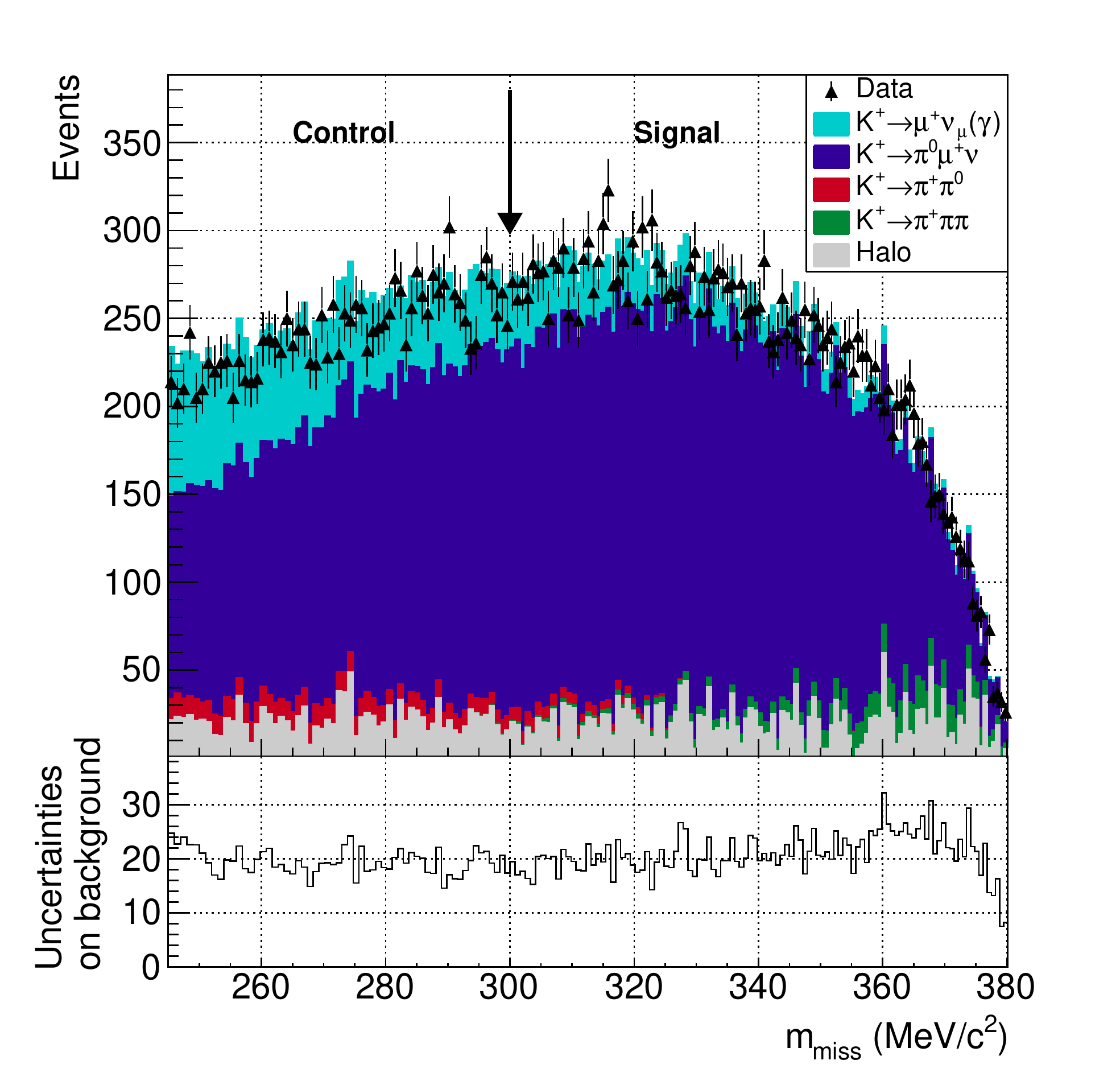}%
\end{minipage}
\put(-160,74){\Large\bf a}
\hfill
\begin{minipage}{0.483\textwidth}
\includegraphics[width=\textwidth]{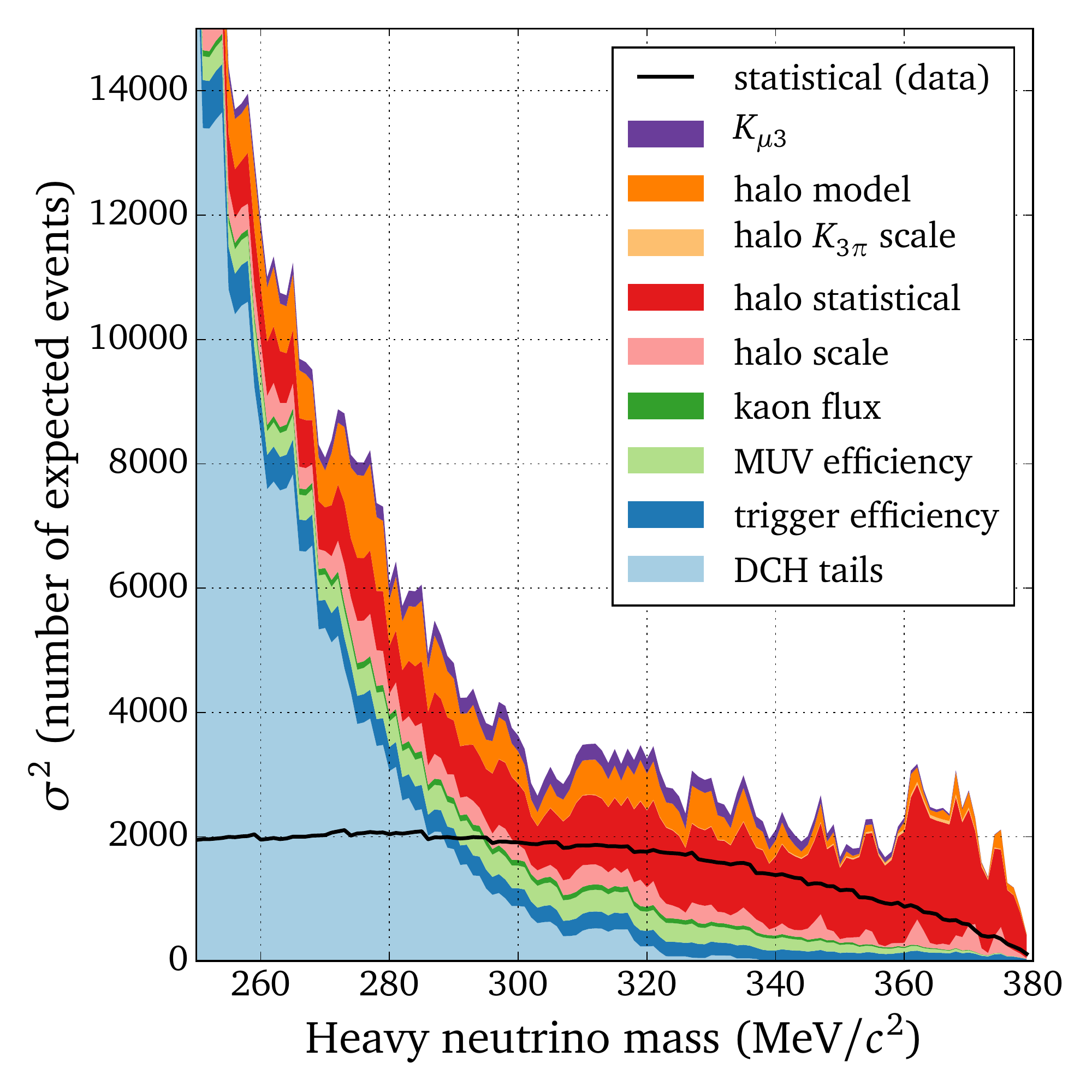}%
\vspace{-0.3cm}
\end{minipage}
\put(-140,74){\Large\bf b}
\caption{a) Missing mass distribution of data and background estimate in the signal and control
regions. Error bars are data statistical errors. The lower plot shows the total uncertainty on the
background estimate. b) Squared uncertainties on the number of expected background events at
each heavy neutrino mass. The coloured bands are the different contributions to the systematic
uncertainties. The black line is the statistical contribution.}\label{fig:kmuN_datasamples}  
\end{figure}

\unboldmath
\label{sec:kmuN_datasamples}
The event selection requires a single positively charged track, reconstructed as a muon.
To suppress the beam halo muons component, five-dimensional cuts in the ($z_{vtx}$, $\theta$, $p$, CDA, $\phi$)
space are applied, where $z_{vtx}$ is the longitudinal position of the reconstructed vertex, $\theta$ is the
angle between the $K^{\pm}$ and $\mu^{\pm}$ direction, and $\phi$ is the azimuthal angle of the muon in the
transverse plane. For the data sample considered, the total number of kaon decays in the fiducial volume, obtained from reconstructed $\kmutwo{+}$ decays, is $N_K = 5.977\times10^7$.
The $\kmutwo{+}$ and other kaon decay background channels are simulated to determine the expected distribution of the reconstructed $M_{miss}$ variable.
The contribution from the beam halo is evaluated using a control data sample, defined as the sample recorded with the $K^-$ beam only.
The missing mass $M_{miss}$ spectrum of data and MC events passing the $\kmutwoN{+}$ selection is shown in Fig.~\ref{fig:kmuN_datasamples}a, together with the total uncertainties on the expected background (systematic and statistical), displayed in the lower plot.
All the contributions to the systematic uncertainties can be seen in Fig.~\ref{fig:kmuN_datasamples}b.
A peak search of the $M_{miss}$ distribution in steps of 1 MeV/$c^2$ is performed.
For each heavy neutrino mass $M_{N_4}$ hypothesis, a window of size $\pm\sigma_M = 120\mbox{ MeV}/c^2 - 0.03 \cdot M_{N_4}$ is used,
corresponding to the resolution on the heavy neutrino mass.

\boldmath
\subsection{Limits on long-living heavy neutrino production}
\begin{figure}[h]
\begin{minipage}{0.52\textwidth}
\includegraphics[width=\textwidth]{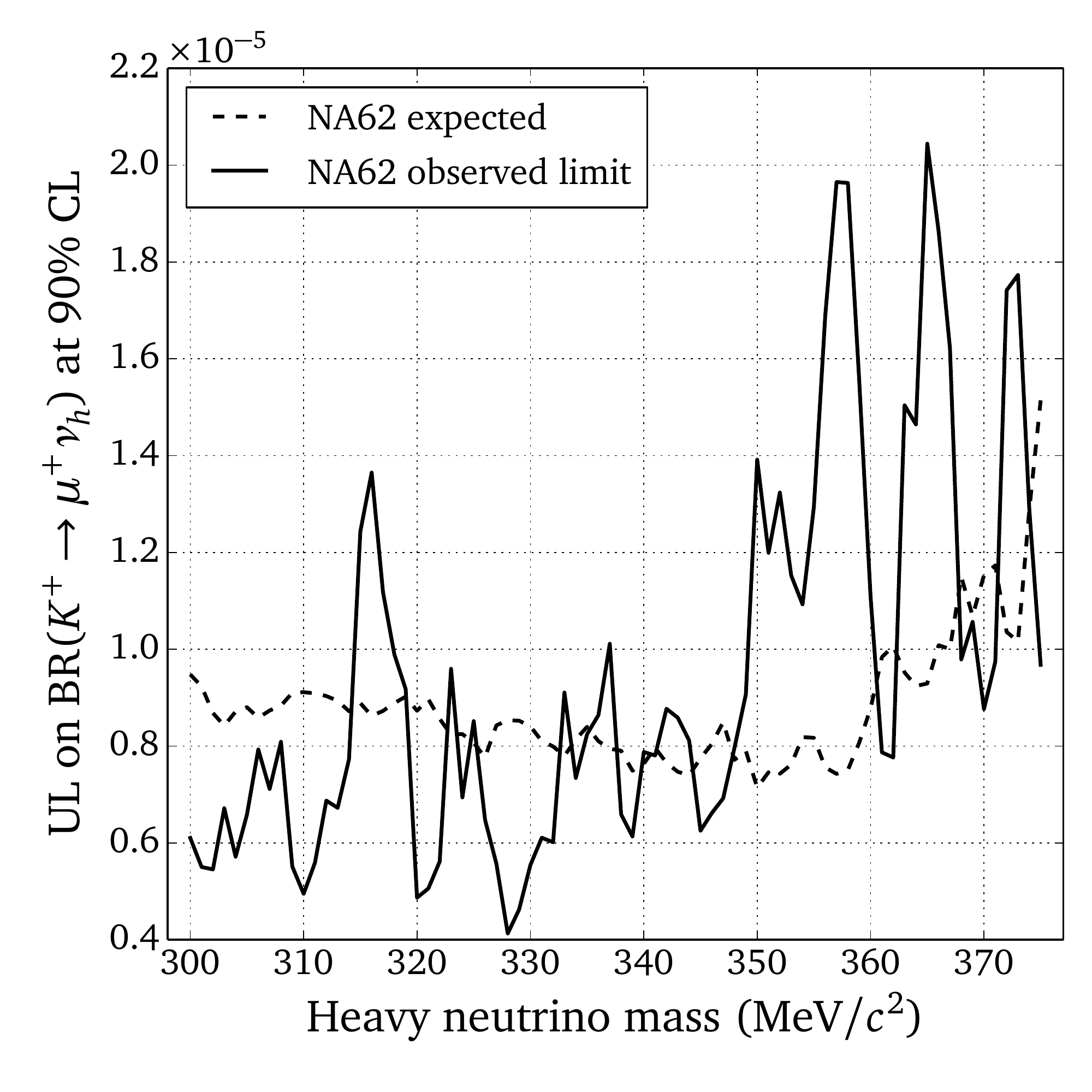}%
\end{minipage}
\put(-23,77){\Large\bf a}
\hfill
\begin{minipage}{0.48\textwidth}
\includegraphics[width=\textwidth]{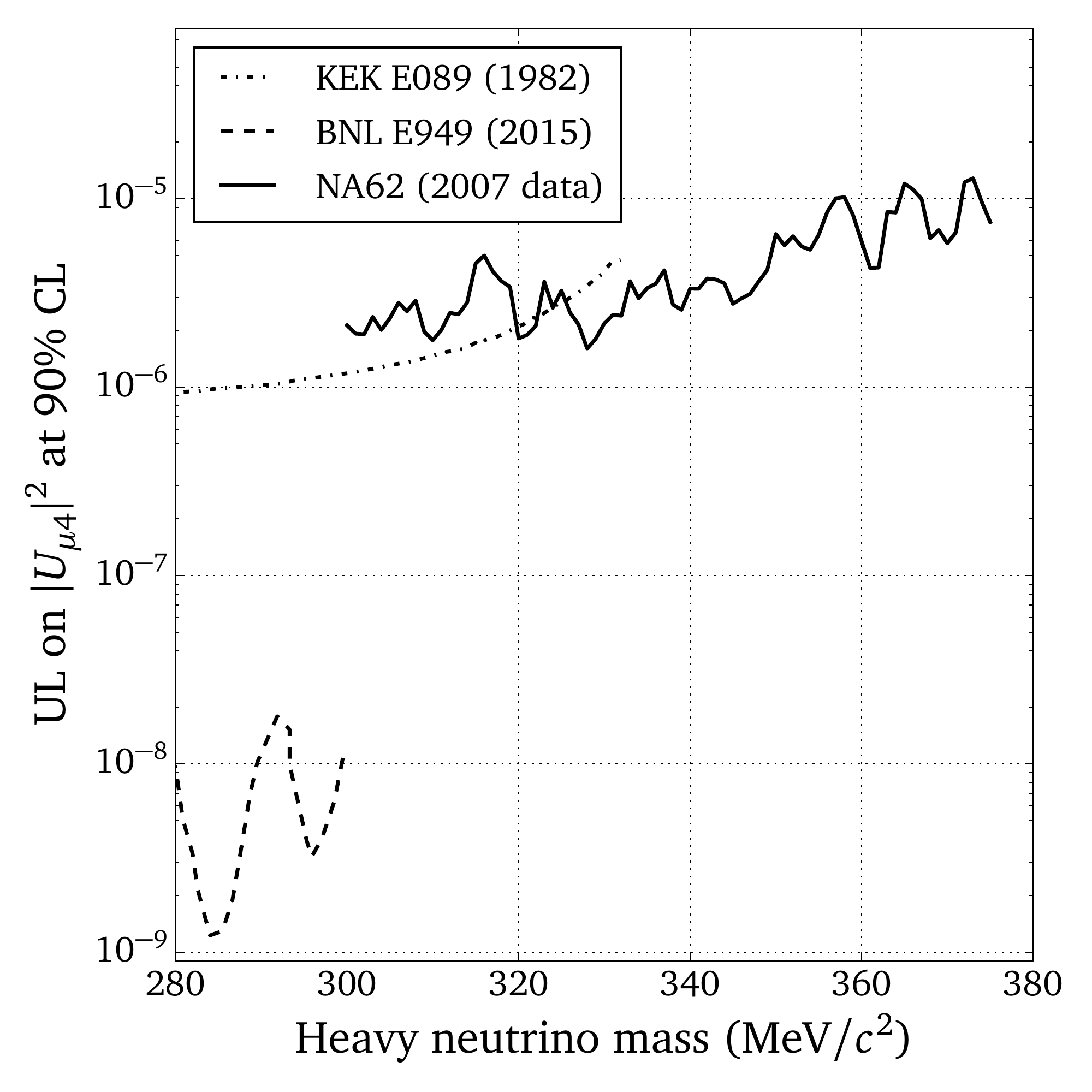}%
\end{minipage}
\put(-23,77){\Large\bf b}
\caption{Obtained ULs at 90\% CL on a) the branching ratio~$\mathcal{B}(\kmutwoN{+})$; b) the magnitude of $|U_{\mu 4}|^2$, as a function of the assumed $N_4$ mass.}\label{fig:KmuN_results}
\end{figure}

\unboldmath
The ULs at 90\% CL on the number of reconstructed $\kmutwoN{+}$ events~$N_{\mu N}^i$ in each mass hypothesis~$M_i$ are computed by applying
the Rolke-Lopez method~\cite{ro01} for the case of a Poisson process in presence of gaussian background.
No signal is observed, as the local significance of the signal in each mass hypothesis never exceeds 3 standard deviations.
In absence of a signal, an UL on~$\mathcal{B}(K^{\pm}\to \mu^{\pm} N_4)$ is obtained for each mass hypothesis~$M_i$, by using the values of the acceptances~$A_{\mu N}(M_i)$ estimated with MC simulations and the UL on the number~$N^i_{\mu N}$ of signal events for such a mass hypothesis:
\begin{equation}
\left.\mathcal{B}(K^{\pm}\to \mu^{\pm}N_4)\right|_{M_i} = \frac{N_{\mu N}^i}{N_K \cdot A_{\mu N}(M_i)}.
\end{equation}
The obtained ULs on~$\mathcal{B}(K^{\pm}\to \mu^{\pm}N_4)$ as a function of the $N_4$ mass are shown in Fig.~\ref{fig:KmuN_results}a.
These limits can be used to constrain the squared magnitude $|U_{\mu4}|^2$ using the relation~\cite{sh81}
$$
|U_{\mu 4}|^2 = \frac{\mathcal{B}(\kmutwoN{+})}{\mathcal{B}(\kmutwo{+})\rho(M_{N_4})},
$$
where $\rho(M_{N_4})$ is a kinematical factor to account for the mass $M_{N_4}$ of the heavy neutrino.
The obtained ULs on $|U_{\mu4}|^2$ as a function of~$M_{N_4}$ are shown in Fig.~\ref{fig:KmuN_results}b.

\section{Conclusions}
Searches for short and long\mbox{-}living heavy neutrinos produced in $\kmutwoN{\pm}$ decays are performed on the large samples of charged kaon decays collected with different trigger conditions by the NA48/2 and NA62-$R_K$ experiments at CERN.
No signals are observed. 

Using the NA48/2 data sample, an UL of $8.6\times10^{-11}$ for $\mathcal{B}(\kpimmws)$ has been established, which improves the best previous limit~\cite{ba11} by more than one order of magnitude. Furthermore, ULs are set on the products~$\mathcal{B}(\kmutwoN{\pm})\mathcal{B}(\Npimuns)$ as functions of the resonance mass and lifetime. These limits are in the $10^{-10}-10^{-9}$ range for heavy neutrino lifetimes below 100~ps.

Using the NA62-$R_K$ data sample, limits on the heavy neutrino production are set at the level of $10^{-5}-10^{-6}$ on the mixing matrix element $|U_{\mu4}|^2$ in the range 300--375~MeV/$c^2$.

\end{document}